\documentclass[12pt]{spieman}  
\usepackage{amsmath,amsfonts,amssymb}
\usepackage{multirow}
\usepackage{graphicx}
\usepackage{setspace}
\usepackage{tocloft}
\usepackage{lineno}
\newcommand{\tcr}{}

\graphicspath{{./}{Figures/}{AP_figs/}}

\title{The Simultaneous Operation of a Controllable Segmented Primary Mirror and Single Conjugate Adaptive Optics System part 2 - Simulated Operation}

\author[a, *]{Benjamin Calvin} 
\author[a]{Michael P. Fitzgerald} 
\author[b, c]{Sam Ragland}
\affil[a]{University of California - Los Angeles, Physics and Astronomy, 430 Portola Plaza, Los Angeles, CA, USA, 90095}
\affil[b]{W. M. Keck Observatory, Kamuela, HI, USA, 96743}
\affil[c]{Large Binocular Telescope, 933 N. Cherry Drive, Tucson, AZ, USA, 85721}

\cftpagenumbersoff{figure}
\cftpagenumbersoff{table} 
\begin{document} 
\maketitle

\begin{abstract}
There are scientific and technological needs to improve the co-phasing of the primary mirrors of segmented telescopes. We have developed a methodology for using the wavefront sensor of an adaptive optics (AO) system to disentangle the phase of a Controllable Segmented Primary mirror (CSP) from the residual phase aberrations to be corrected by the rest of the AO system. 
We show simulations of the Keck-II AO system where we simultaneously control the CSP surface and a single-conjugate AO system using a pyramid wavefront sensor in a low-bandwidth closed loop, resulting in significantly decreased aberrations on the primary and improvements in the performance of the AO system. Analysis of on-sky telemetry data from the Keck-II AO system validates these simulations and shows that accurate measurements of the primary phase can be made in the presence of real atmospheric turbulence. Ultimately, this work suggests that it is feasible to simultaneously monitor and control the CSP with an AO wavefront sensor while in closed-loop operation and that doing so may significantly improve the performance of the AO system. 

\end{abstract}

\keywords{astronomical instrumentation: adaptive optics, astronomical instrumentation: high angular resolution, techniques: high angular resolution}

{\noindent \footnotesize\textbf{*}Benjamin Calvin,  \linkable{bcalvin@astro.ucla.edu}, \linkable{ben.a.calv@gmail.com} }


\section{Introduction} \label{sec: intro}

Segmented telescopes are important to both current and future optical and near-IR astronomy, and the alignment of the segmented primary mirror plays an essential role in enabling these telescopes to achieve their scientific goals. 
As was discussed in part one of this two part paper, most segmented telescopes maintain the alignment of their primary mirrors by using capacitance-based edge sensors to detect pairwise motion between consecutive segments \cite{1990SPIE.1236.1038C}. However, investigations into the phasing of the Keck primary mirrors have suggested that the performance of these control systems are not \tcr{generally meeting the evolving needs for extreme wavefront control.} 
\cite{2018SPIE10700E..1DR, 2022SPIE12182E..09R, 2023ApOpt..62.5982C}.

In part one, we introduced a methodology to use the wavefront sensor of the adaptive optics (AO) systems in these large segmented telescopes as a supplementary alignment sensor for the Controllable Segmented Primary mirror (CSP) in simultaneous operation with AO correction \cite{2025calvin..A}. 
\tcr{This methodology isolates the component of the wavefront sensor signal that is uniquely controllable by the CSP and removes the component that is controllable by the deformable mirror (DM).} 
\tcr{We generate this spatially filtered CSP interaction matrix as} 
\begin{equation}
    A_{\text{iCSP}} = A_{\text{CSP}} - A_{\text{DM}} \times A_{\text{DM}}^{-1} \times A_{\text{CSP}},
  \label{eqn:CSP_ind_matrix}
\end{equation}
where $A_\text{iCSP}$, $A_\text{CSP}$, and $A_\text{DM}$ indicate the interaction matrices of \tcr{the wavefront sensor responses to the CSP in the independent space, in the mixed space, and to the DM respectively.} 
$A_\text{DM}^{-1}$ denotes the pseudo-inverse of the DM interaction matrix, dubbed the DM control matrix. 
We once again highlight that despite being redeveloped for our specific context, this methodology is mathematically equivalent to the distributed modal command methodology presented by Ref. \citenum{2007ApOpt..46.4329C}.

While this mathematical framework does not necessarily assume a specific AO architecture, we chose to model a simulated AO system after the Keck-II AO \tcr{(KAO)} system using the IR pyramid wavefront sensor in NGS mode\tcr{,} 
as this PyWFS was recently added to KAO as part of the 
Keck Planetary Imager and Characterizer (KPIC) instrument \cite{2016SPIE.9909E..0DM, 2020JATIS...6c9003B, 2021JATIS...7c5006D}. 
In imaging mode, KPIC consists of many upgrades to the KAO system to improve image quality on NIRC-2, including the PyWFS, high throughput optics, and new AO control algorithms to improve contrast at low spatial separations \cite{2019SPIE11117E..0WJ, 2021SPIE11823E..1FV}. In characterization mode, KPIC couples light into a single-mode fiber to provide NIRSPEC with a stable line spread function from a highly spatially filtered target and includes many hardware upgrades to maximize the coupling efficiency into the fiber \cite{2019SPIE11117E..0UP}. These include, again, the same upgrades to the KAO system, along with a Zernike wavefront sensor \cite{2011SPIE.8126E..0FW}, apodizers \cite{2021PASP..133b4503C}, a high-order deformable mirror, vortex coronagraphs, and an improved tracking system \cite{2022SPIE12184E..1WE}. 
While added as part of the KPIC instrument, the PyWFS has been made available for AO operation as a single-conjugate AO system in the infrared. As an SCAO system, all of the available information in the wavefront phase is encoded into the singular output of the PyWFS, including both the phase aberrations from atmospheric turbulence and from misalignments in the CSP surface. 

In part one, we showed from simulation that the AO architecture of KAO could yield measurements of the \tcr{average} CSP \tcr{mode in the independent space} to a precision limit of approximately \tcr{$\sim 16$\,nm} 
rms across the pupil when observing a target of brightness \tcr{$m_H < 10$} 
or brighter for a total integration time of 30\,seconds. 
In this work, we simulate the simultaneous closed-loop operation of a CSP and KAO using the PyWFS, and we develop methodologies to generally characterize the CSP in closed loop. 
In \S\ref{sec:CSP_sim_methods}, we describe a test simulation to demonstrate the simultaneous control of the the AO system and the CSP on different timescales both with and without the presence of atmospheric turbulence. 
In \S\ref{sec:CSP_results}, we present the results of these simulations. 
In \S\ref{sec:On-sky}, we apply this approach to on-sky PyWFS telemetry data from Keck II. 
These topics are then discussed in \S\ref{sec:CSP_discussion} before reviewing the key conclusions from both part one and this paper in \S\ref{sec:conclusion}.

\section{Methods for Characterizing a CSP+AO System} \label{sec:CSP_sim_methods}
\tcr{Our main goal is to determine whether the alignment of the CSP can continue to be measured to high accuracy in closed-loop operation.} 
Equation \ref{eqn:CSP_ind_matrix} --- introduced in part one --- 
was 
presented in a manner separated from the specific architecture of the CSP+AO system. 
However, we will now primarily focus on a simulation of a combined CSP+AO system using KAO as the primary AO system. 
This will allow us to implement specific tests to gauge the capabilities and advantages of the CSP method. 
In \S\ref{sec:sim_setup}, we setup our simulation of KAO. In \S\ref{sec:setup_noATM}, we describe the control laws for the CSP and our metrics for assessing the performance of the CSP in the case of zero incident phase aberrations. The results of these tests are presented in \S\ref{sec:results_noATM}. In \S\ref{sec:ATM_char}, we introduce incident phase aberrations from atmospheric turbulence and comment on how they impact the CSP methodology. The results of these tests are presented in \S\ref{sec:results_ATM}.

\subsection{Simulation Setup}\label{sec:sim_setup}

We have created a simulation modeled after the KAO system using HCIPy \cite{por2018hcipy}. 
We've constructed a 10\,m primary aperture consisting of 36 hexagonal mirror segments with a 2.4\,m circular central obstruction acting as the secondary mirror. \tcr{We have included} 
a deformable mirror with 21$\times$21 Gaussian actuators that leads into a PyWFS, modulated at $5\lambda /\text{D}$, with 88$\times$88 pixels, which is exposed to an incoming wavefront and generates images of four sub-apertures of the pupil that can be used to recreate an estimation of the wavefront. In this simulation, we use $\lambda = 1600$\,nm as both the wavefront sensing wavelength and science camera wavelength. 
\tcr{We simulate a $m_H = 0$ point source guide star so that we may operate in the high SNR regime of this proof-of-concept study. As a result, we do not include any sources of noise at this stage.}

While 
the KAO system uses these sub-apertures to calculate the slopes of the wavefront in the $x$ and $y$ directions to generate commands for the deformable mirror \cite{2020JATIS...6c9003B}, we have opted to directly use the image output in a return-to-reference operation, as described in Ref. \citenum{2016Optic...3.1440F}, which defines the signal as the difference between the image output and a reference image from a \tcr{static} 
wavefront. 
Both techniques retain different advantages \cite{2022SPIE12185E..1BB}, but there is a larger domain in the form of more pixels in the PyWFS images, meaning the independent subspace is larger in the images method compared to the slopes method. This facilitates calculation of accurate iCSP commands using images, compared to using slopes. 
In Sec. \ref{sec:On-sky}, we show some complementary results from using on-sky telemetry data in the slopes domain that portray similar findings to the results that we will show in Sec. \ref{sec:CSP_results}, indicating that the choice of wavefront sensor domain is flexible when monitoring the primary phase. %

We can use the simulation of the KAO system to build interaction matrices for the DM and CSP in mixed space, and then use Eq. \ref{eqn:CSP_ind_matrix} to \tcr{create} 
the independent, spatially filtered CSP interaction matrix. By taking the pseudo-inverse of this interaction matrix, we get an independent control matrix for the CSP that will now take a given output from the PyWFS and determine the commands for the CSP that will recreate the component of the output independent of the DM controllable space through the independent PyWFS responses. 
The 
commands generated by this filtered control matrix are not influenced by the state of the deformable mirror within the linear range of the PyWFS system. An inherent caveat of the linear separation used in Eq. \ref{eqn:CSP_ind_matrix} is the failure to handle non-linear effects that may blend the controllable spaces of the two optics 
-- the implications of which, we will discuss further in Sec. \ref{sec:CSP_discussion}.

\begin{figure}[t] 
\centering
\includegraphics[width = 0.99\textwidth]{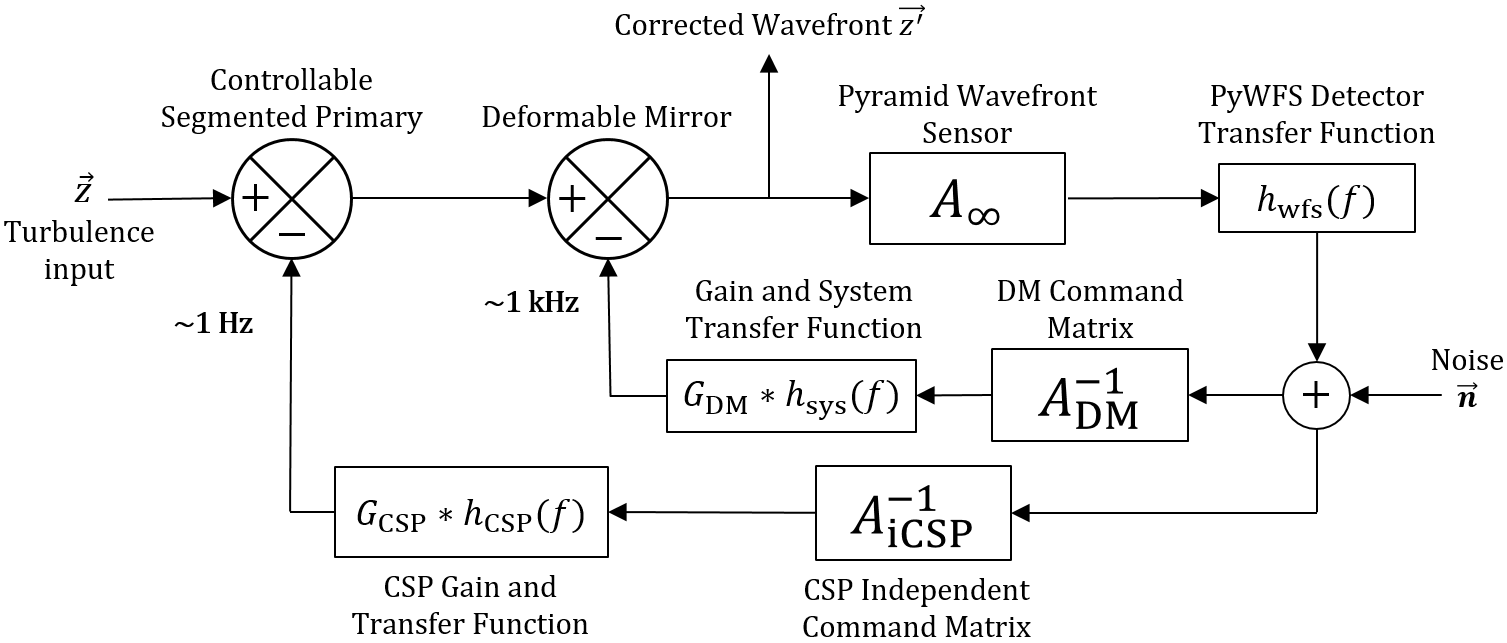}
\caption{Schematic representation of our combined CSP+AO control loop, modeled after Ref. \citenum{1994A&A...291..337G} Figure 1. Here, the original AO loop remains unchanged while the CSP is sensed using auxiliary information from the wavefront sensor. \label{fig:block_diagram}} 
\end{figure}

To demonstrate the ability to autonomously flatten the CSP in closed loop while the AO system is simultaneously \tcr{also} in closed loop, 
we must develop a functional form of the control scheme for commanding the primary mirror that ensures the operability and non-invasiveness of the CSP+AO system. Beginning in a state of misalignment, the CSP induces phase aberrations in the wavefront. 
Then, the combined CSP+AO system operates simultaneously, where the DM is instructed to correct the wavefront unobstructed at a high temporal cadence and the CSP integrates commands from the independent space through the duration of its slower integration time. At the completion of the CSP integration, the \tcr{primary segments are} 
updated according to the wavefront reconstruction and gain. We visually describe this update step in Fig. \ref{fig:block_diagram}, using the same structure to represent the AO system as in Ref. \citenum{1994A&A...291..337G}. 
It is mathematically represented as: 
\begin{equation}
    \vec{z}_{j} = \vec{z}_{j-1} - G_{\text{CSP}} \times \left(A^{-1}_{\text{iCSP}} \times \vec{b}_{\text{av}}\right),
    \label{eqn: CSP_upd}
\end{equation}
where $\vec{z}_{j}$ is the set of commands for the CSP modes at step $j$, $G_{\text{CSP}}$ is the gain of the CSP, $A^{-1}_{\text{iCSP}}$ is the independent command matrix from Eq. \ref{eqn:CSP_ind_matrix}, and $\vec{b}_{\text{av}}$ is the average output from the PyWFS in the mixed wavefront space between steps $j-1$ and $j$. \tcr{The labels $j$ and $j-1$ denote the relative timesteps of the CSP control system.}

We also define the speeds and control laws that govern the CSP+AO system. For the AO system, we implement a rudimentary proportional control law with a high gain and a low leakage 
at 1\,kHz operation. \tcr{The precise type of used control law will be an important decision in an implementation of this methodology for facility operation, but in this controlled simulation, it is inessential. } 
As the purpose of this work is to demonstrate how the CSP can be incorporated into a CSP+AO system without affecting the behavior of the AO loop, as depicted in Fig. \ref{fig:block_diagram}, 
\tcr{the control parameters} are set to be consistent across every simulation. 
Meanwhile, as the CSP already cannot operate at the same temporal bandwidth as the AO system, 
we have the freedom to not only assign the CSP its own gain and leakage parameters, but also its own effective integration time. 

However, %
the leakage term cannot be considered in quite the same way as it is for a DM. Basic AO control is often implemented in the form of a "leaky integrator", 
where the leakage term allows the DM to return to some pre-referenced flat wavefront state. Due to the infrequent calibrations for mirror phasing and the mis-calibrations that occur within the primary mirror edge sensor systems, the true flat state of the primary mirror segments cannot be assumed to be known. Thus, the inclusion of leakage into the CSP control law would risk increasing the phase aberrations present across the primary mirror. 
Since this would be counter-productive, we set the CSP leakage to 0\% for the remainder of the work. Thus, the two free parameters that determine the closed-loop bandwidth of the CSP control system are the integration time and the gain. 

\subsection{Characterizing the CSP with Zero Random Wavefront Error}\label{sec:setup_noATM}

In this subsection, we introduce metrics to evaluate how well we can \tcr{monitor and correct the CSP phase in closed loop} 
and the methods for measuring these metrics. To demonstrate the operability of the combined CSP+AO closed-loop system, we first execute our simulation in an environment without the presence of atmospheric turbulence. 
In HCIPy, we're able to propagate a wavefront at a chosen wavelength through the different optics making up the CSP+AO system, allowing us to measure properties of the wavefront at different locations along the optical path. This property allows us to make measurements to quantify the performance of the CSP+AO system in-situ (e.g. accurately isolating the surface deviations of the primary from other optical aberrations or measuring the Strehl ratio following the DM with zero NCPAs).

There are two critical aspects for characterizing this system. As we are implementing the CSP methodology to correct and control aberrations present from the primary mirror, we must track a level of the flatness of the primary. We do this by calculating the RMS surface deviation of the primary mirror from its known simulated flat reference shape. %
This metric directly tracks the accuracy of our CSP commands. Additionally, we want to present these results in a manner that is immediately applicable to observers. As such, we also track the Strehl ratio of the PSF, 
recorded in the final image plane following the CSP+AO system with zero NCPAs assumed in the rest of the optical path. We will record the evolution of these metrics through the operation of the CSP+AO system in closed loop. 

Here, we seek to demonstrate the behavior of the CSP in closed-loop control, so we set our simulations to run for a pre-determined number of update steps to capture both the corrective and stable behaviors of the CSP in closed-loop operation. 
In the case where there are no phase aberrations originating from a turbulent atmosphere, we applied sets of Gaussian noise equally to the piston, tip, and tilt modes of the CSP segments to generate an approximately $\sim 120$\,nm rms initial 
\tcr{surface} aberration across the pupil\tcr{. We use random Gaussian noise to induce the aberrations on the primary in an attempt to recreate the spatial distribution of CSP modes observed in the initial misaligned piston maps measured in Ref. \citenum{2022SPIE12182E..09R}, despite not recreating the periodic piston errors investigated in Ref. \citenum{2018SPIE10700E..1DR}. The} 
CSP+AO system then autonomously corrected its shape back to proper alignment. In Sec. \ref{sec:results_noATM}, we show and comment on the results of these runs\tcr{, which will validate the behavior of the CSP+AO system under ideal conditions}. 

\subsection{Characterizing the CSP with a Simulated Atmosphere} \label{sec:ATM_char}
Large ground-based telescopes will always be affected by phase aberrations from the atmosphere. After validating the CSP+AO system in a simulation in ideal conditions, our goal will be to show that we can still execute closed-loop operation of the CSP and develop expectations for biasing effects in the presence of turbulence. 


For this, we must consider all of the \tcr{relevant} areas of wavefront space that phase aberrations exist in. 
While atmospheric turbulence injects phase aberrations into all wavefront spaces, \tcr{to impact the behavior of the CSP,} the phase aberrations must either be within the independent CSP space or within a space that is aliased into the independent CSP space --- both of which are outside of the controllable space of the deformable mirror, by definition\tcr{.} 
Thus, the atmospheric phase aberrations that are relevant to the CSP must be uncorrectable by AO. 

Atmospheric turbulence also injects phase aberrations on all relevant timescales. While typically, the dynamical timescale of the atmosphere is significantly shorter than the response time of the primary, there will be portions of signal in the independent CSP space from the atmosphere that evolve on timescales similar to the CSP integration time. It is not our intent to command the CSP in response to this signal, but rather to maintain the alignment of the primary mirror. These atmospheric turbulence signals will bias the CSP commands and have the potential to act as a limit to the accuracy of the CSP. By increasing the integration time of the CSP, 
we \tcr{can} 
decrease 
the random error induced from the atmospheric residuals. It is beyond the scope of this work to characterize exactly how the properties of the atmospheric turbulence residuals impact the qualities of the measurements of the independent CSP signal. This topic was lightly addressed in Ref. \citenum{2023SPIE12680E..0CC}. 

HCIPy generates simulations of atmospheric layers utilizing the method presented in Ref. \citenum{2006OExpr..14..988A} to make phase screens that obey Kolmogorov statistics and dynamically evolve through frozen-flow behavior. On timescales longer than $\sim 1$ second \cite{1999SPIE.3762..225S, 2014MNRAS.440.1925G, 2022SPIE12185E..83C}
, frozen-flow does not generally 
%
%
represent the behavior of atmospheric evolution. In an effort to mimic boiling and other non-linear effects seen in real turbulence \tcr{while preserving computational performance}, we use \tcr{a composition of opposite-velocity pairs of atmospheric layers at different altitudes that combine to a $C_n^2$ profile, total Fried parameter, $r_0$, and decorrelation time, $t_{\text{\tcr{dec}}}$, that can be modeled after the atmosphere at any specific location.} 
We go on to control the CSP at a rate with timescales significantly longer than the timescale of the boiling and non-linear effects of the atmosphere, meaning the exact details of these effects will be less impactful in our investigation. However, this remains a limitation of the simulation, and any conclusions we reach must be independently validated, ideally with on-sky observations.

Knowing that the atmosphere will generate errors in the commands for the CSP, we require a new metric to determine how well the free parameters of the CSP can mitigate the errors in commands induced by the atmosphere. This can be described as the "robustness" of the system, which tracks the average variance in the amplitudes of each CSP mode during closed-loop operation when the accuracy of the CSP is limited by atmospheric turbulence. 

Additionally, as the presence of atmospheric turbulence will cause temporal variability in the Strehl ratio, we calculate the change in the Strehl ratio by comparing the Strehl in the final image plane while running the CSP+AO system to the Strehl ratio with only the AO system. Because these atmospheric profiles can be repeatedly generated in simulation, this comparison shows only the impact of the CSP on the Strehl ratio. 

Therefore, to systematically measure the impact that the effective integration time and gain have on the performance of the CSP in the presence of atmospheric turbulence, we track the following three performance metrics: 
\begin{itemize}
    \item Average change in Strehl ratio, comparing the average Strehl to a simulation that does not update the primary from the initial misaligned state,
    \item Average primary flatness, measuring the average surface deviations in alignment of the primary mirror, and
    \item General primary robustness, measuring the standard deviation in the flatness of the primary across update steps.
\end{itemize}
We measure these quantities for a series of gain values from \tcr{30\% 
to 80\%} 
and with effective integration times ranging from 5 seconds to 11 seconds. \tcr{This range of integration times would not typically be sufficient to eliminate atmospheric residuals from the measurement of the CSP modes, but they reveal trends in system behavior within the limitations of our computational resources that will hold true for longer integration times. The behavior of the atmosphere on longer integration times will ultimately need to be investigated with further on-sky study.} 
We report on the results of these simulations in Sec. \ref{sec:results_ATM}.

\section{Simulation Results} \label{sec:CSP_results}
Here, we summarize our findings describing the 
utility of using the PyWFS to simultaneously drive both the CSP and the AO system. 

\subsection{Simultaneous Control with Zero Random Wavefront Error} \label{sec:results_noATM}
In this section, we present the results of running the CSP+AO system simultaneously in closed loop without the influence of a turbulent atmosphere. In this case, the only source of phase perturbation is either from the CSP or from the deformable mirror. We recorded 25 iterations of 
closed-loop \tcr{tests with} 
differing random initial surface aberrations and present the average behavior in this section.

\begin{figure}[t]
\centering
\includegraphics[width = 0.80\textwidth]{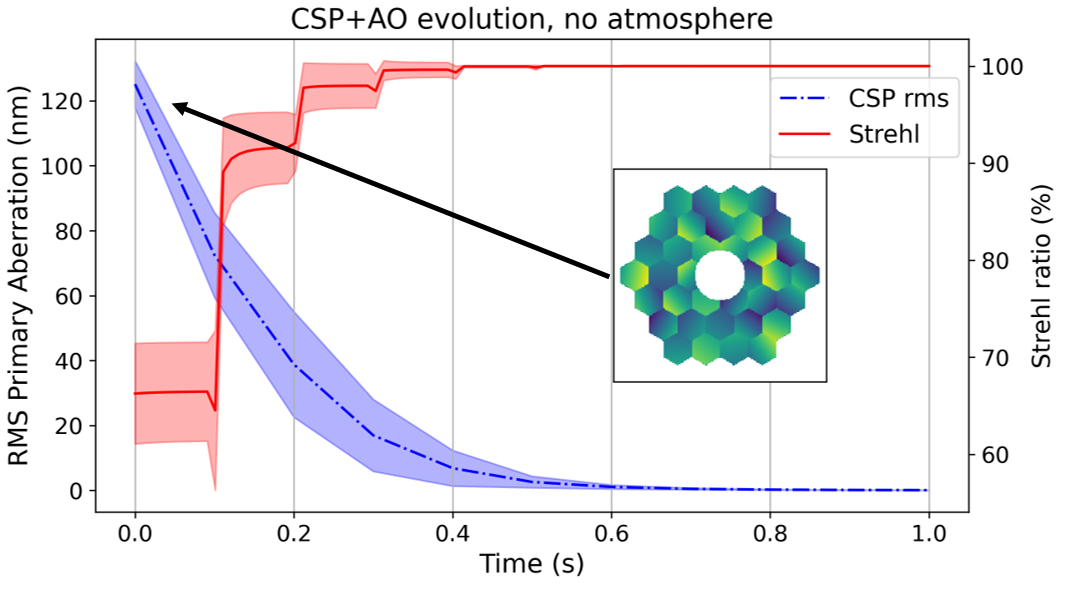}
\caption{The result of the no-atmosphere closed-loop CSP+AO simulations. In blue, the rms surface evolution curve shows a monotonic decrease in the phase aberrations present on exclusively the primary. In red, the Strehl evolution curve shows the change in behavior of the wavefront following the CSP+AO system. This plot suggests a causal relationship between flattening the primary and achieving a high Strehl ratio. \label{fig:CSP_noATM_res}}
\end{figure}

The set of initial 120\,nm rms surface aberrations were generated on the primary through piston, tip, and tilt modes, \tcr{as seen in the inset of Fig. \ref{fig:CSP_noATM_res}}. 
Since this test is a proof of concept, the CSP was set to run at an nonphysical 10\,Hz and the AO system at 100\,Hz to flatten the wavefront aberrated by the CSP surface perturbation. 
Additionally, the gain of the CSP system was set to 95\%. The rms \tcr{misalignment} of the surface and the Strehl ratio of the final wavefront are recorded as the system updates, generating Fig. \ref{fig:CSP_noATM_res}. The blue curve of the figure records the evolution of the surface aberration on the primary. The left y-axis is the rms misalignment of the primary in nanometers, and the shared x-axis denotes the time in the simulation in seconds. In this no-atmosphere case, this simulation was run for 10 CSP steps, where we see that the primary \tcr{surface} was flattened monotonically to less than 1\,nm rms. 

Meanwhile, the co-plotted red curve records the evolution of the Strehl ratio in \tcr{\textit{H}-band at} the final image plane following the CSP+AO system. In the 10 AO update steps in between CSP steps where the wavefront incoming from the CSP is unchanging, the AO system is able to correct the phase perturbations controllable by the DM, minimizing the combined system's perturbations. 
The Strehl ratio curve shows how the phase perturbations generated by the primary will still affect the final wavefront quality, despite the corrections from the AO system. 
We refer to the post-AO Strehl ratio, for an aberrated wavefront that cannot be perfectly corrected by the deformable mirror, as the maximum achievable Strehl ratio for that wavefront. Within four updates to the CSP, 
the maximum achievable Strehl ratio has increased from \tcr{an initial} 55\% to over 97\% in the absence of atmospheric turbulence. 

There are noticeable decreases in the Strehl ratio of the system that occur at some update steps. This is because there is currently no compensatory command sent to the deformable mirror when the CSP is updated in our simulation. As we did not implement a non-zero response time to CSP commands in our simulation, the phase aberration that the AO system is attempting to correct instantaneously changes, causing a decrease in Strehl prior to the AO system responding. This effect would not be observed in an on-sky implementation, as the response time of the primary mirror segments is longer than the response time of the AO system \cite{2004ApOpt..43.1223C}. Applying an update to the CSP shape in a real system will appear as a correctable continuous process to the AO system.

\subsection{Simultaneous Control with Atmosphere} \label{sec:results_ATM}

In this section, we present the results of running the CSP+AO system simultaneously in closed loop, just as before in Sec. \ref{sec:results_noATM}, but with the presence of a turbulent atmosphere while varying the CSP control parameters as described in Sec. \ref{sec:ATM_char}. 
We seek to make observations for how the integration time and gain of the CSP affect the performance of the combined CSP+AO system.

\begin{figure}[t]
\centering
\includegraphics[width = 0.80\textwidth]{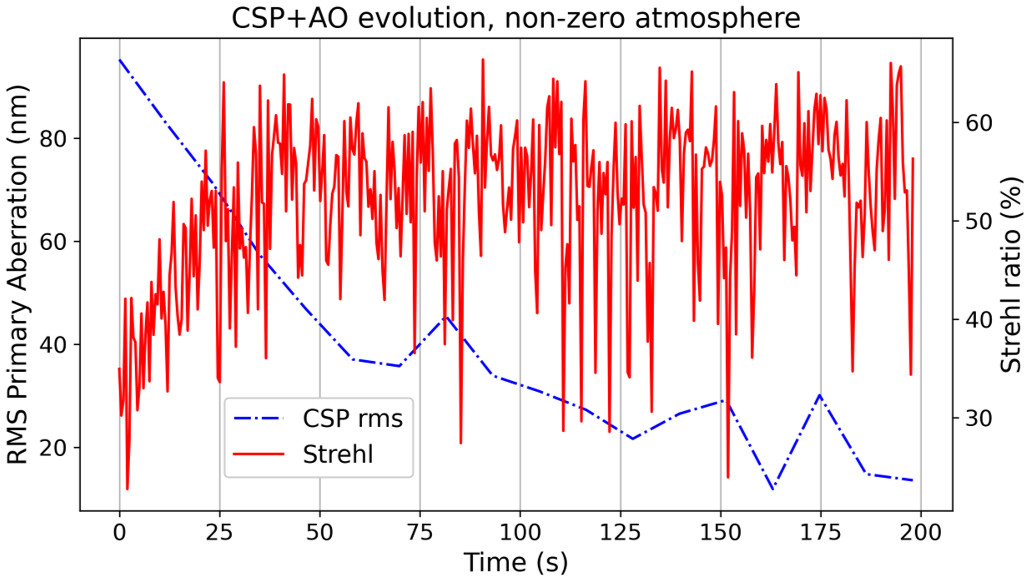}
\caption{Results of a closed-loop CSP+AO simulation with a simulated atmosphere present. This run used a gain of \tcr{40\%} 
and an integration time of 11 seconds. The rms surface evolution curve (blue) shows that the initial phase perturbation 
is flattened to about $\sim 20$\,nm RMS, where the residual phase from the atmosphere is the dominant term influencing the primary alignment. The Strehl evolution curve (red) shows an increase in the Strehl following the CSP+AO system, coincident with correction of the primary. This uptick shows an approximate $\sim 25\%$ increase in Strehl in \textit{H}-band. \label{fig:CSP_longInt_res}} 
\end{figure}

In Fig. \ref{fig:CSP_longInt_res}, we present the results from an example simulation that used an integration time for the CSP of 11 seconds and a gain of \tcr{40\%} 
with the AO system operating at 1\,kHz to correct an atmosphere with a Fried parameter of \tcr{$r_{0} = 22$\,cm, corresponding to 0.45 arcsec seeing at 500\,nm, and $t_{\text{dec}} = 27$\,ms. The typical seeing on Mauna Kea is expected to be approximately 0.75\,arcsec with a \tcr{decorrelation} time of 5\,ms \cite{2009PASP..121..384S}. While the simulated conditions are optimistic compared to the typical conditions for the Keck telescopes, they are still realistic conditions that allow us to use a rudimentary integrator control loop without needing to fully optimize the AO system to achieve our project goals. } 

This simulation, which ran for 18 update steps, was given an initial perturbation placed on the CSP \tcr{surface} of 90\,nm rms. From the blue RMS alignment curve, we can see that 
the primary does not converge to the same level of alignment as in the case without turbulence, but is instead limited by the presence of the atmosphere. 
For this example simulation, the lower limit of the alignment --- the average primary flatness --- is roughly \tcr{20\,nm rms.} 
When the alignment was atmosphere-limited, 
the standard deviation of the rms \tcr{surface} alignment in time is \tcr{7.5\,nm,} 
which describes our robustness metric introduced in \S\ref{sec:ATM_char}. 

From the red Strehl ratio curve of Fig. \ref{fig:CSP_longInt_res}, we see the initial Strehl ratio for the combined system is approximately \tcr{37\%} 
during closed-loop AO while the primary is misaligned. 
After the CSP loop is closed, the Strehl ratio increases to approximately \tcr{$\sim 55$\%.} 
We can observe that the increase in the Strehl ratio coincides with the initial improvement in CSP alignment, much like was observed in Fig. \ref{fig:CSP_noATM_res}. 
In this plot, the atmosphere generates high frequency temporal variations in the Strehl measurements, indicating that the response time of the AO system is significantly faster than the dozens of seconds over which the Strehl ratio increased at the beginning of the simulation. Therefore, we are confident that the initial misalignment of primary is the limiting factor for the Strehl ratio at the beginning of the simulation
, and that within 5 update steps of the CSP, the Strehl ratio was only limited by the post-AO atmospheric residuals and no longer affected by the phase of the CSP. %

While the results in Fig. \ref{fig:CSP_longInt_res} demonstrate the closed-loop control is possible in the presence of turbulence, we must collect more than one data point in order to properly build expectations of the behaviors of the CSP+AO system. 
To this extent, we've systematically investigated the effect of the selected free parameters of the CSP, the integration time and gain, on how the CSP+AO system responds to atmospheric turbulence. For a given atmosphere \tcr{with the same properties as the simulation in Fig. \ref{fig:CSP_longInt_res} in each run,} 
we vary the gain from \tcr{30\% to 80\%} 
and the integration time from 5 seconds to 11 seconds and track the metrics described in Sec. \ref{sec:ATM_char}. 

\begin{figure}[t]
\centering
\includegraphics[width = 0.95\textwidth]{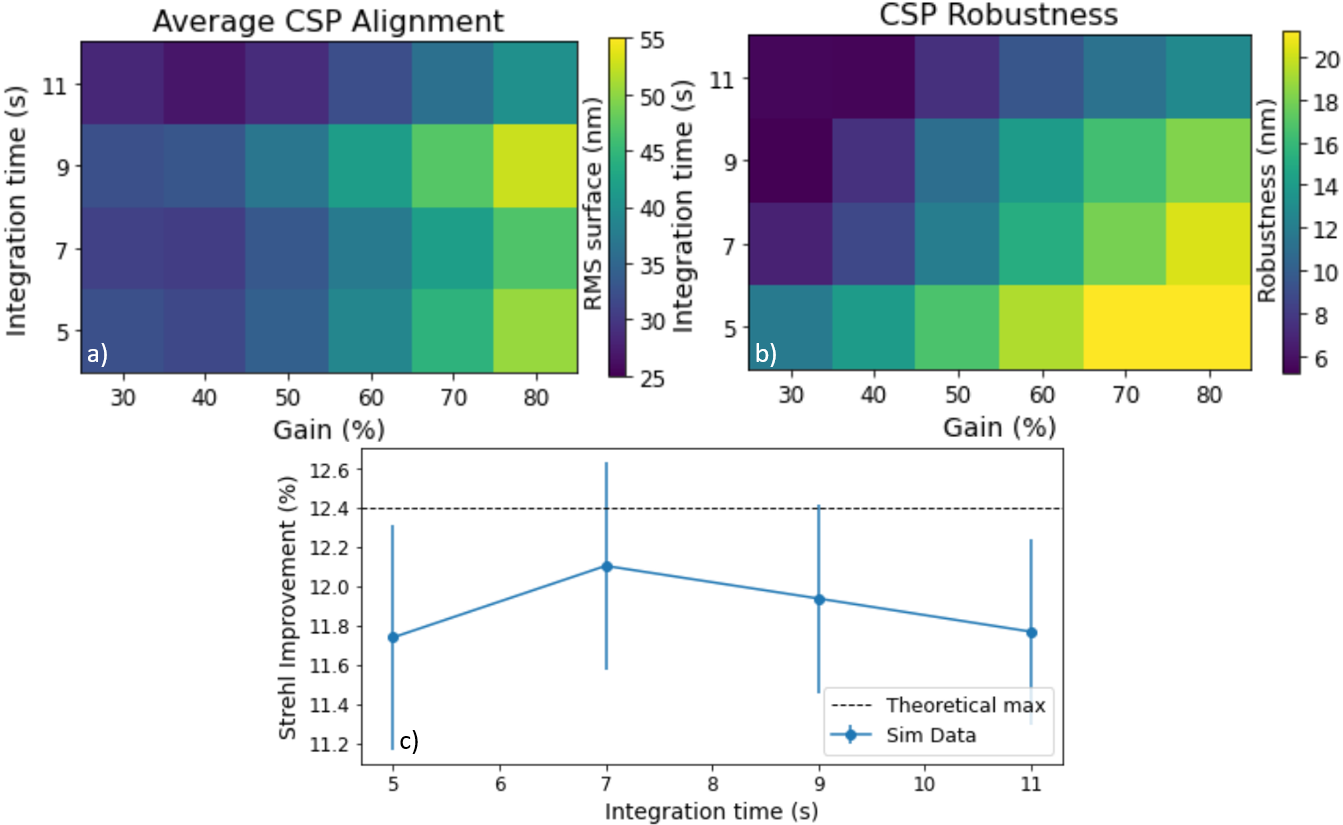}
\caption{Demonstration of how the CSP control parameters affect the alignment and robustness of the primary and the Strehl following the CSP+AO system. a) We show the average converged alignment of the CSP. The alignment \tcr{generally improves by decreasing the gain and by increasing the integration time.} 
b) We show the robustness of the converged alignment of the CSP, which improves monotonically both by decreasing gain and increasing integration time. The performance of the 11 second integration time and 40\% gain outperforms the requirements from Ref. \citenum{1994SPIE.2199..105C}. c) We show the Strehl improvement in comparison to the maximum achievable improvement, which shows that in every case, the Strehl ratio of the CSP increases to near the theoretical maximum. \label{fig:grid_search}}%
\end{figure}

In Fig. \ref{fig:grid_search}, we present the findings of this investigation for the flatness, robustness, and Strehl metrics. In Fig. \ref{fig:grid_search}a, we show how the average converged flatness of the CSP changes in response to the gain and integration time. The figure shows that as we either decrease the gain or increase the integration time, we observe a near monotonic improvement of the flatness
. In subplot b, we show 
the robustness of this alignment between update steps and how it varies across the free parameters. 
Together, the two subplots depict the broad behavior of the CSP with different parameters. There are many combinations of gain and integration time that produce similar CSP flatness results, but general trends in the data suggest that the use of a lower gain generates a better robustness, despite changes in the integration time. 
In concert, these two subplots give an estimation for what control parameters will be appropriate for a given observation. 

In Fig.~\ref{fig:grid_search} subplot c), we show how the Strehl ratio improved in these simulations. The Strehl ratio improvement was independent of the gain, so we present this metric only against the integration time. The error bars are the temporal average of the Strehl improvement measurements, made by comparing Strehl from the CSP+AO system to those of the AO-only simulation. \tcr{We also plot the theoretical maximum improvement of the Strehl ratio, which is found by comparing the average Strehl ratios from a simulation with a flat primary to a simulation that doesn't update the initial primary aberration.} 
From this subplot, we see that, by updating the CSP in simultaneous operation with the AO system, nearly equivalent improvements to the Strehl are achieved, without regard for the specific behavior of the primary. The 5-second integration time case yields the smallest improvement to Strehl, but still reaches within 10\% of the improvement that we would expect to see from a perfect primary alignment. %
The implications that this investigation has for other observing scenarios is discussed in Sec. \ref{sec:CSP_discussion}.

\section{Open-Loop Tests with On-Sky Telemetry Data} \label{sec:On-sky} 
To further validate both our methodology and our simulation results, we conducted an open-loop response test to assess this method using on-sky data. 
In February 2022, an investigation was conducted using Keck-II to explore methods for improving the performance of the Keck telescopes' primary mirror technology \cite{2022SPIE12185E..0YR}. In this investigation, individual segments from the Keck-II primary were pistoned relative to the remainder of the pupil to form a characteristic "L-mode", using 3 segments in an L-shaped arrangement, at amplitudes of 50, 100, 200, and 400\,nm. For this paper, we have acquired telemetry data from closed-loop operation of the PyWFS at each of these L-mode amplitudes, as well as at a baseline 0\,nm amplitude reference. 
This grants us access to PyWFS outputs corresponding to known segment offsets. While these data were not taken directly with our CSP+AO methodology in mind, they still allow us to test our CSP method on real on-sky data and compare our results to known inputs. In this section, we investigate the linearity of the PyWFS response to the L-mode at varying amplitudes, compare the structure of the PyWFS response to the L-modes to simulation, and test recovering measurements of the CSP phase using an interaction matrix generated from this dataset. Our goal in this section is to show that we can use on-sky AO telemetry data to accurately recover known phase aberrations placed on the primary mirror, which will provide further confidence in the methodology we've introduced in this project. 

\begin{figure}[t]
\centering
\includegraphics[width = 0.98\textwidth]{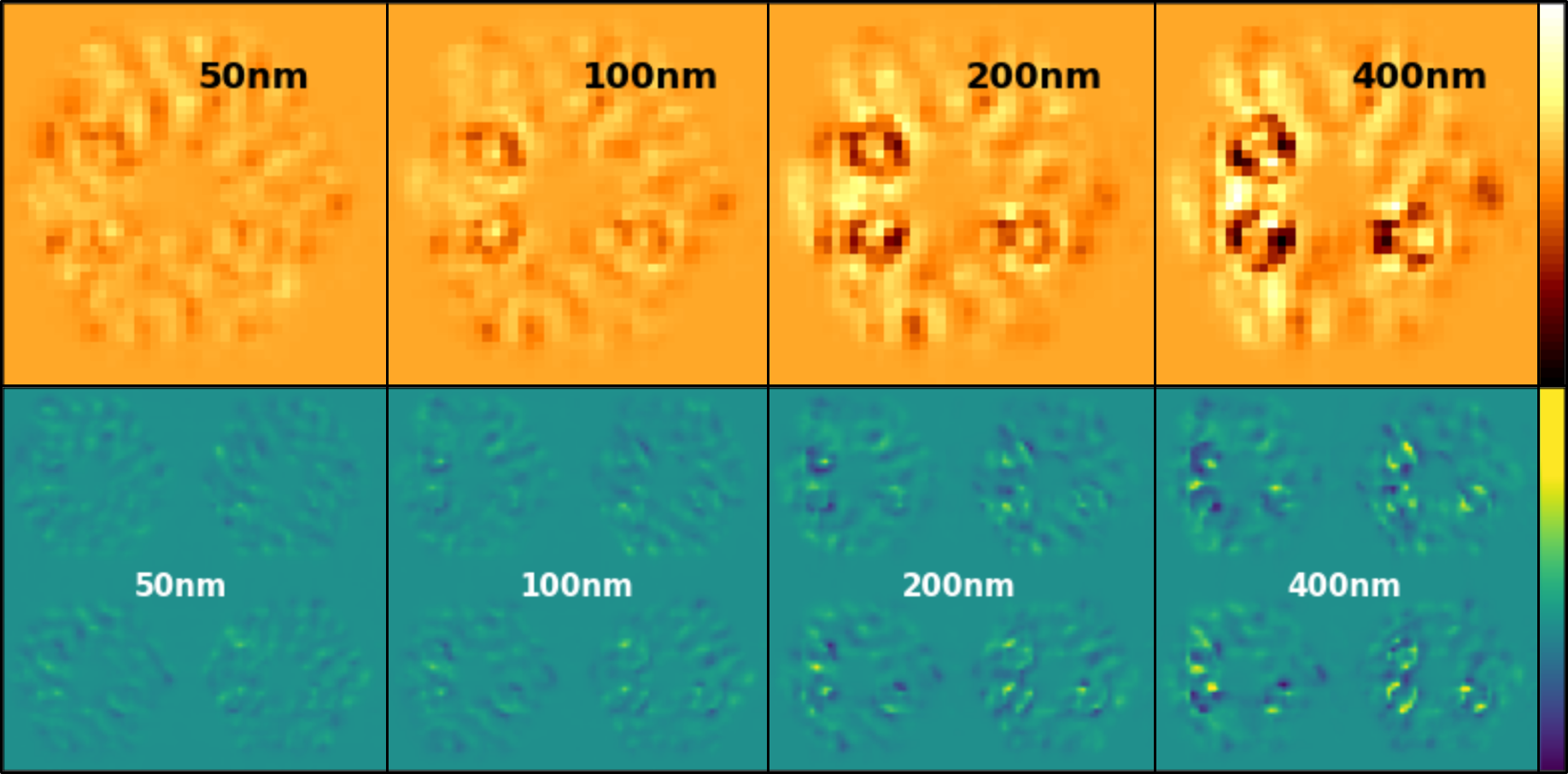}
\caption{The referenced on-sky measurements of the L-modes at varying amplitudes. Top row: the x-slopes of the L-modes. Bottom row: the PyWFS images of the L-modes. In both cases, we can see that the 400\,nm mode is not within the linear range of the PyWFS. \label{fig:sky_data}}
\end{figure}

\tcr{The atmospheric characteristics on Mauna Kea are recorded in the Mauna Kea Weather Center seeing archive.\footnote{\url{http://mkwc.ifa.hawaii.edu/current/seeing/index.cgi}} Unfortunately, the MASS/DIMM data from the observation night, February 14, 2022, was not recorded. The previous night had an average seeing of $1.4$\,arcsec and the following night had an average seeing of $0.7$\,arcsec. The observation night had temperature, pressure, and humidity statistics more similar to the following night, while the wind speed was more similar to the preceding night. We are unable to acquire further information about the atmospheric statistics of the on-sky data. }

The PyWFS was operating in high-speed mode at 1\,kHz and captured one minute of data for each L-mode amplitude. This gives 60,000 data points for each measured amplitude. As the correlation time of the Mauna Kea atmosphere is typically estimated to be 5\,ms, the minute of telemetry data corresponds to 12,000 correlation times. Knowing that the sensing target from this investigation had an \textit{H}-band brightness of 5 or brighter \cite{2022SPIE12185E..0YR}, the Precision Limit --- defined in Sec. 3 of part one --- 
predicts that the uncertainty in CSP commands from noise in this telemetry data to be well less than 5\,nm, the step resolution for the Keck primary mirror segment actuators \cite{1994SPIE.2199..105C}. 

The telemetry data contain both dark-subtracted images from the PyWFS, processed $x$ and $y$ slopes from the PyWFS output, and the interaction matrix of the DM. In both the images and the slopes, due to static phase offsets present in the rest of the observatory, 
each dataset must be subtracted by the reference data to observe the effect of inducing the L-mode on the PyWFS. 
The average reference-subtracted image and $x$-slope are shown in Fig. \ref{fig:sky_data} for each L-mode amplitude. In the top row of the figure, we show the referenced $x$-slope of the pupil, and in the bottom row we show the referenced images. We note that the amount of signal at the location of the pistoned segments increases by a factor of $\sim 1.6$ between the 50\,nm and 100\,nm frames and by $\sim 2.1$ from the 100\,nm to 200\,nm frames. However, when comparing the 200\,nm frames to the 400\,nm, the signal changes by a factor of approximately $\sim 1.02$\tcr{. This suggests that 200\,nm of piston is near the limit of the linear range of this system, but this conclusion cannot be accepted without skepticism, due to the sparse amplitude sampling of this dataset.} 

\begin{figure}[t]
\centering
\includegraphics[width = 0.80\textwidth]{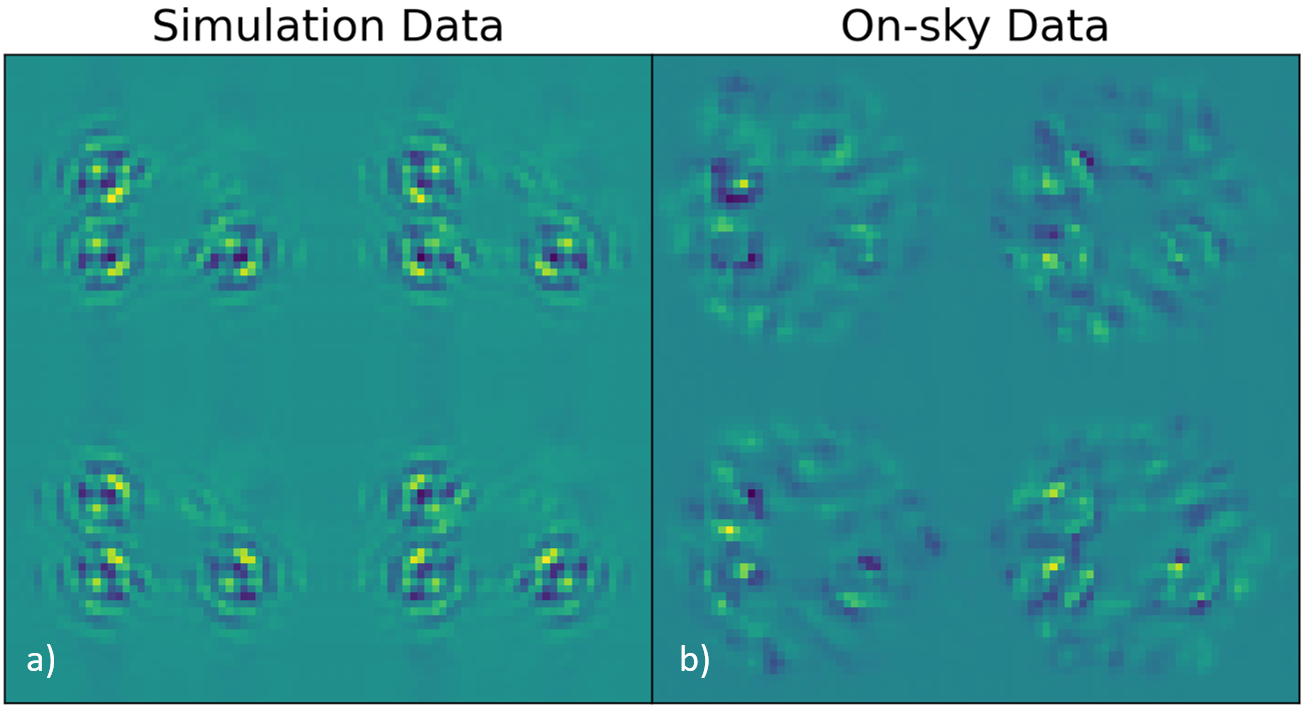}
\caption{Comparison between the PyWFS response to the L-mode at 200\,nm in the independent wavefront sensor space. While sharing similar characteristics, there are features that appear in the on-sky data (b) that are not present in simulation (a). \label{fig:sky_compare}}
\end{figure}

In Fig. \ref{fig:sky_compare}, we compare the 200\,nm L-mode image with what we would expect in the independent space from simulation. \tcr{In this simulation, since we do not control the CSP in closed loop and because we do not know the precise atmospheric characteristics of the observation, we attempt to match the median atmospheric parameters of the Maunakea atmosphere. We set the Fried parameter to $r_0 = 12$\,cm, which corresponds to a seeing of $0.84$\,arcsec at $\lambda=500$\,nm, and a decorrelation time of $t_\text{dec}=0.018$\,s. We then subtract the average PyWFS image across 60\,seconds of integration with the L-mode aberrated at 200\,nm from the average PyWFS image at the reference amplitude. }
This figure shows clear similarities in the structure between the simulation and sky-data (e.g. the size of the signal in the pupil images and the orientation of the bright spots in both). 
However, the simulated and measured signals exhibit substantial differences due to intricacies in accurately modeling the KAO lightpath that would fall beyond the scope of this project. These differences could be from NCPAs, conjugation issues, imperfect optics, differences in the pyramid optical layout, etc. 
Despite the similarities building confidence in our methodology used in simulation, the differences between simulation and reality preclude us from easily using the simulated CSP interaction matrices to analyze the on-sky data. 

We can, however, use a sample of the sky-data to create a limited interaction matrix of the CSP. This was done by isolating the regions of the pupil containing each individual segment to act as an approximate zonal basis. We used the 200\,nm dataset to build this interaction matrix due to its large signal compared to atmospheric residuals while still remaining within the linear range of the PyWFS. 
Then, by using the DM interaction matrix, which is provided in the telemetry data, we can use Eq. \ref{eqn:CSP_ind_matrix} with the limited CSP interaction matrix to generate the independent CSP interaction and control matrices.

The KAO system operates using the $x$ and $y$ slopes calculated from the PyWFS images, as opposed to the images domain which we use in our simulations
. As the provided DM interaction matrix uses the slopes domain, 
we must build a\tcr{n independent} CSP interaction matrix in this domain as well. 
Due to the fact that the domain choice is non-critical and was not specified \tcr{earlier,} 
we are free to make this change without expecting any significant negative impact to our wavefront recovery.

\begin{table}
    \centering
    \begin{tabular}{|c|c|c|c|c|c|c|}
        \hline
        Known & Seg. 1 &  & Seg. 2 &  & Seg. 3 &  \\
        Seg. Amp. & Recall & Seg. 1 Error & Recall & Seg. 2 Error & Recall & Seg. 3 Error \\
        ($nm$) & ($nm$) & ($\%$) & ($nm$) & ($\%$) & ($nm$) & ($\%$) \\ [1ex]
        \hline\hline
        \rule[-1ex]{0pt}{3.5ex}  50  & 55.6  & 11.2  & 59.2  & 18.4  & 51.5  & 3.0  \\
        \hline
        \rule[-1ex]{0pt}{3.5ex}  100 & 94.7  & 5.3   & 96.7  & 3.3   & 111.8 & 11.8 \\
        \hline
        \rule[-1ex]{0pt}{3.5ex}  200 & 199.4 & 0.3   & 200.7 & 0.35  & 201.3 & 0.65 \\
        \hline
        \rule[-1ex]{0pt}{3.5ex}  400 & 214.1 & 46.8  & 201.2 & 49.7  & 199.9 & 50.0 \\ [0.5ex]
        \hline
    \end{tabular}
    \caption{Recoveries of the On-sky iCSP Amplitudes}
    \label{tbl:On-sky}
\end{table}

With this, we can make measurements from the datasets at the other known amplitudes to observe how accurately we can recover the segment piston positions on independent data. 
In Table \ref{tbl:On-sky}, we list the recoveries of the three segments' piston amplitude from each dataset. As the interaction matrix was built from the 200 nm dataset, we naturally recover this amplitude with the least amount of error. For the 50 and 100\,nm amplitude L-modes, our iCSP method consistently measured the offset to within an accuracy of 10\,nm using the slopes domain. 
\tcr{The 400\,nm amplitude L-mode was recovered to only approximately $\sim 50\%$ accuracy, further suggesting that} 
this amplitude extends beyond the linear range of the PyWFS. However, despite the large amount of error in the measurement of the 400 nm piston, the incorrect measurement will still \tcr{return} 
the segments closer to alignment, demonstrating the potential of closed-loop operation to still be effective in this non-linear regime.



\section{Discussion} \label{sec:CSP_discussion}
From \tcr{our} 
analysis, it is clear that there is potential for using the wavefront sensor of an AO system to correct for misalignment errors between segments of a segmented primary mirror. We go on to discuss the implications that these results have on segmented telescope operation, and comment on steps toward implementing this method on-sky.

\subsection{Maximum achievable Strehl} \label{sec:max_strehl}
From the simulations in the zero random wavefront error case, shown in Fig. \ref{fig:CSP_noATM_res}, without considering any atmospheric effects on the wavefront, there are two important takeaways. The first is the demonstration that, by observing the independent wavefront space, the CSP method can accurately determine the position and alignment of each segment and autonomously flatten the sensed aberrations in the primary. This means that the only barrier to operation in this simulated framework is amassing a sufficient signal over all sources of uncertainty to ensure accurate measurements of the primary phase. 
The second key takeaway comes from how the Strehl ratio of the wavefront following the CSP+AO system is initially $<$100\%. As this simulation takes place in an environment with no other source of wavefront error, this limited performance is a result of the AO system being incapable of fully correcting the misalignments of the primary. However, 
by updating the CSP with a few of our measurements made in the independent space, the rms \tcr{surface} misalignment from the primary is reduced to $<$10\,nm rms and asymptotically approaches perfect alignment in every iteration of the simulation. From there, the AO system is able to flatten the remaining phase aberration and achieve a Strehl ratio of $>$99\%. 


\tcr{Figure \ref{fig:CSP_noATM_res} provides both a measure of input wavefront error and the resulting Strehl ratio. We would expect the relationship between the two to be related 
to the Mar\'echal approximation\cite{1952iga..book.....M}, 
\begin{equation}
    S \sim \text{exp}\left(- \sigma_\text{post}^2\right), \quad \sigma^2_\text{post} = k\cdot \sigma_\text{in}^2,
    \label{eqn:modified_mare}
\end{equation} 
where $\sigma_\text{post}$ is the phase variance following the CSP+AO system, $\sigma_\text{in}$ is the initial CSP wavefront error, and $k$ is assumed to be some constant factor by which the DM corrects the initial CSP wavefront error. Careful observation of Fig. \ref{fig:CSP_noATM_res} reveals that this is not the case.} 
The rms variance in wavefront phase does not evolve linearly with a changing magnitude of CSP aberration \tcr{because} 
the PyWFS 
non-linearly \tcr{responds} 
to the independent signal \tcr{at} 
large CSP amplitudes --- affecting the DM's response to the CSP modes. 
By varying the amplitude of CSP aberrations, 
we can determine the level of phase aberration from the CSP at which the \tcr{non-linear effects begin to bias the correction the DM applies to the CSP misalignment. } 

\begin{figure}[t]
\centering
\includegraphics[width = 0.75\textwidth]{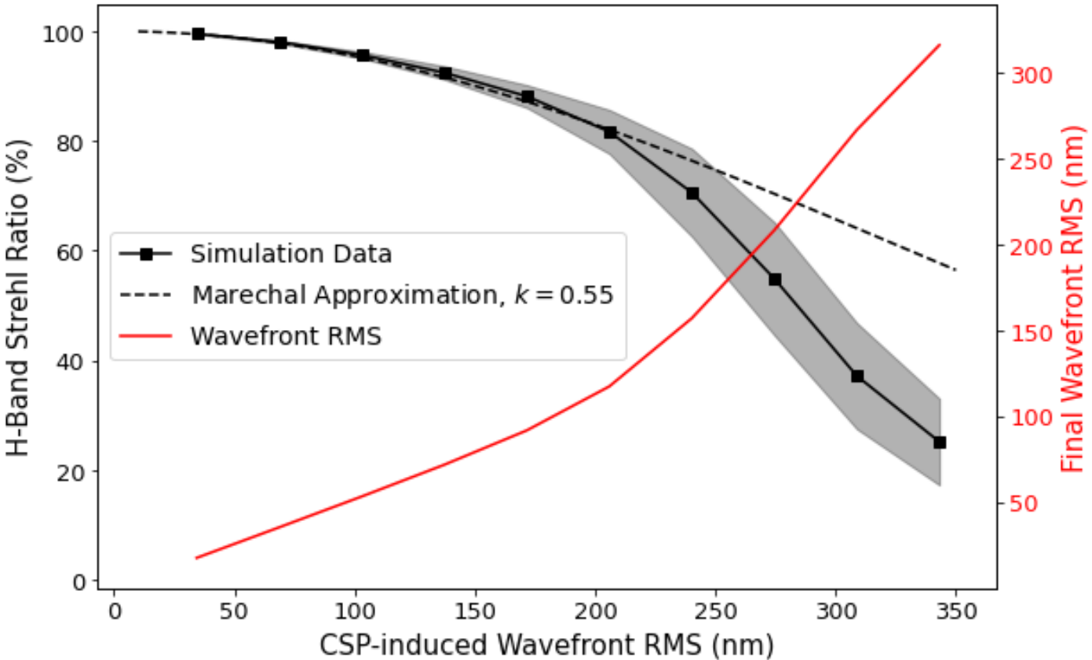}
\caption{A visual representation of the relationship between the phase aberrations from the CSP, and how the AO system responds. In black, we show the evolution and uncertainty of the Strehl ratio as it changes with the aberrated CSP surfaces, as well as what would be expected from the \tcr{modified} Mar\'echal approximation\tcr{, Eq. \ref{eqn:modified_mare}.} In red, we show how the flatness of the wavefront following the CSP+DM system evolves in response to the CSP aberrations. There is a clear break in the wavefront flatness curve that corresponds to when the DM is unable to approximate the CSP, which coincides with the knee in the Strehl ratio plot. \label{fig:rms_track}}
\end{figure}

In Fig. \ref{fig:rms_track}, we expand on this. We vary a series of surface aberrations of the CSP from 20 to 170\,nm rms each -- inducing 40 to 340\,nm rms aberrations of optical path difference -- and allow the AO system to correct the aberrated wavefronts. \tcr{These surface aberrations are generated in the same manner as those from the previous simulations, shown in Fig. \ref{fig:CSP_noATM_res}.} We then plot the output wavefront rms phase aberration in red and, in black, we plot the \tcr{\textit{H}-band} Strehl ratio in the final image plane 
for that level of input phase aberration. We also plot the expected Strehl ratio from the \tcr{modified} Mar\'echal approximation \tcr{from Eq. \ref{eqn:modified_mare} with $k=0.55$} in a dashed black line. Through 100\,nm rms of surface aberrations on the CSP, the observed simulation Strehl ratios \tcr{matches well to the Mar\'echal approximation, due to the AO correction of the CSP modes.} 
However, beyond this magnitude of surface aberration, the measured Strehl ratio begins to decrease, coinciding with an increase in the post-CSP+AO phase aberration. This is coincident with the peak-to-valley magnitude of the input phase aberrations entering the non-linear range of the PyWFS, further suggesting that the DM is correcting CSP aberrations until static non-linear effects causes errors in the reconstruction of the CSP surface.

From Fig. \ref{fig:rms_track}, if we consider a segmented primary mirror with \tcr{120\,nm} 
rms misalignment, which would induce a \tcr{240\,nm} 
rms phase aberration in the wavefront, as found in Ref. \citenum{2022SPIE12182E..09R}, we would only expect a maximum Strehl ratio of approximately $\sim 70\%$ in the \textit{H}-band or $\sim 90\%$ in the \textit{K}-band, prior to any interference from the atmosphere. From this state, we could expect to see a $\sim 10\%$ increase in science PSF core flux 
by flattening the primary. This is consistent with the results seen by the on-sky work using the Zernike wavefront sensor to align the Keck primary mirror \cite{2024ApJ...967..171S}. \tcr{Since the phase aberrations controlled by the CSP in our simulation are defined as the aberrations that cannot be controlled by the deformable mirror of the AO system, sharing consistent results suggests that the improvements to the Strehl ratio achieved in Ref. \citenum{2024ApJ...967..171S} are the result of correcting phase errors that originate on the primary mirror, rather residuals from the AO system. }

\tcr{Further investigation shows that for arrangements of the CSP modes that overlap with the controllable space of the DM, as would typically be expected as the residual surface errors from the edge sensing systems \cite{1998SPIE.3352..307T, 2004ApOpt..43.1223C}, this deviation from Eq. \ref{eqn:modified_mare} occurs at much higher incident phase aberrations. This further suggests that applying updates to the shape of the primary mirror that result in increases to the final Strehl ratio do so by correcting phase errors that cannot be corrected by the AO system. 
Since the piston error maps recovered by Refs. \citenum{2018SPIE10700E..1DR, 2022SPIE12182E..09R, 2024ApJ...967..171S} more closely resemble the uncorrelated random errors that we investigate in this paper than those of the edge sensor residuals, we use Fig. \ref{fig:rms_track} in Sec. \ref{sec:CSP_sim_discuss} to further validate the results we presented in Sec. \ref{sec:CSP_results}.
While the findings from this specific figure will eventually require further validation with physical optics, these results suggest that,} with proper calibration and sensing, our method for maintaining the alignment of the primary mirror can provide substantial improvements on the Strehl ratio and wavefront stability through the AO system.

\subsection{Simulated Operation} \label{sec:CSP_sim_discuss}
Figure \ref{fig:CSP_longInt_res} presents an example of using simultaneous CSP+AO control in the presence of atmospheric turbulence. In blue, the figure shows that the improvement to the rms alignment of the primary mirror. Following the correction of the initial misalignment, the CSP maintained the alignment \tcr{to a comparable level to the requirements of the Keck M1 control system\cite{1994SPIE.2199..105C}. }
Despite the CSP flatness being limited by the atmosphere in the later stages of the simulation, the corrective behavior of the CSP in the early stages is very similar to the case with no contribution from the atmosphere, suggesting consistent behavior by the CSP+AO system when CSP phase aberrations are large in comparison to atmospheric residuals.

The purpose of the simulations including the atmospheric turbulence models beyond this individual example is to investigate how the free parameters of the CSP system, the integration time and the gain, mitigate the random and systematic errors in commands imparted by the atmosphere. The variations of these two parameters were compared against the three metrics of Strehl increase, average achievable alignment, and alignment robustness. From Fig. \ref{fig:grid_search} a) and b), we determined that using a low gain is helpful in reducing the residual atmospheric phase errors to achieve an aligned and stable primary. The presented data would suggest that a lower gain parameter is always preferable. However, a gain of 0\% in the CSP control law would not lead to improvements in the alignment of the primary, as the primary would not update its segment positions. This implies that there is some ideal gain parameter that is high enough to allow the control system to adapt the primary shape to measured misalignments on an appropriate timescale, while not being so high as to be significantly influenced by the atmospheric phase residuals that persist on the timescale of the integration time. 

This ideal gain during typical operation will likely depend on the integration time of the CSP and the decorrelation time of the atmosphere. Reference \citenum{1994SPIE.2199..105C} sets a parameter similar to how we use gain to $10\%$ in the M1 control system based on the edge sensor measurements --- which can serve as an initial expectation for our ideal gain. This could not be thoroughly investigated in simulation both due to computation constraints for the length of simulation necessary to implement a gain that low and due to the inaccuracies in the behavior of simulated atmospheric profiles. Further analysis will require additional data as well as an initial on-sky implementation of our control system, which will be saved for future investigation. 
Otherwise, the main outcome from these subplots is that, despite the presence of atmospheric interference, we remain able to accurately sense and flatten the primary to a stable state. This is reflected by the initial on-sky data that we presented in Sec. \ref{sec:On-sky}, which help to validate the findings from our simulation.

From Fig. \ref{fig:grid_search} c), we also observed that for the specific set of atmospheric properties for this closed-loop investigation, almost every combination of control parameters yielded near theoretical maximum improvement to the Strehl, based on our expectations from Fig. \ref{fig:rms_track}, despite the fact that the varying control parameters resulted in different levels of CSP alignment and robustness. 
The initial perturbation of the primary in this study has a misalignment of 85\,nm rms, inducing an aberration in the wavefront of 170\,nm rms. From Fig. \ref{fig:rms_track}, we can expect that the DM can only bring the original wavefront to a maximum achievable \tcr{\textit{H}-band} Strehl ratio of about $\sim 90\%$, meaning that, by flattening the CSP, we can expect about a $\sim 10\%$ increase in Strehl, which is what we observe from the simulated data. 
Phase aberrations with a magnitude of $\sim 170$\,nm \tcr{rms OPD,} 
as were imparted by the CSP in our closed-loop investigation, are near the turnover point of the Strehl curve where the DM is no longer capable of effectively removing the phase aberrations from the CSP. By flattening the primary even down to a magnitude of $\sim 40$\,nm RMS, imparting 80\,nm RMS onto the wavefront, we successfully return the wavefront to the \tcr{controllable domain of the DM} 
and recover near theoretical maximum improvement to the Strehl. In Sec. \ref{sec:results_ATM}, we noted that the improvement to the Strehl ratio was independent of the CSP gain parameter, while the gain parameter observably affected both the alignment and the robustness of the CSP. Therefore, despite the time-varying errors in the primary alignment induced by atmospheric bias, these errors 
did not affect the improvement to the Strehl ratio. This implies that these atmosphere-induced errors primarily fall within the controllable space of the DM, and that the regimes where the atmosphere limits the accuracy of the CSP may not impact the stability of the science PSF as would otherwise be expected. As mentioned before, the limitations of the HCIPy in realistically evolving atmospheric profiles 
mean that these potential implications must be verified from on-sky data.

Knowing the physical mechanism that sets the limit to the performance of the CSP+AO system, whether atmospheric fitting residuals on the DM, misalignments of the primary, bandwidth errors, high-order aliasing, or any other source of error in the AO system, helps to inform the ways we can use the CSP to improve AO performance. 
By making use of our CSP methodology, we can effectively remove semi-static phase aberrations on the primary mirror from the wavefront error budget \cite{2022SPIE12185E..0YR, 2023aoel.confE...8W} which would remove one source of slowly evolving speckles in the Keck PSF. Motivated by our accurate on-sky measurements of individual CSP modes, presented in \S\ref{sec:On-sky}, we will begin to address the stability of the PSF along with the robustness, repeatability, sensitivity, resilience to atmospheric effects, and other aspects of the CSP in future on-sky investigations. 

\subsection{Looking Toward Implementation}

Though there is clear potential in our approach for improving the alignment of the primary mirror in segmented telescopes, there are a number of assumptions and considerations that have been 
used within this project that may complicate the operation of a CSP+AO system in a more realistic implementation. 
These include assumptions on the control scheme of the CSP, regarding pupil drift and rotation on the detector and phase wrapping ambiguities, and assumptions on the CSP calibration, regarding the time investment and accuracy in building the CSP interaction matrix. In this section, we bring attention to some known limitations of this control scheme that arise from focusing on the fundamental elements of our work, but we also highlight work done by others that address these limitations.

The modulated PyWFS has a well-documented limitation to sensing discontinuities in phase that span longer than a wave \cite{2000SPIE.4007..416E, 2022JATIS...8b1515H, 2022A&A...658A..49B}, and while many techniques have been developed to be able to sense and correct $n\lambda$ phase wrappings and non-linearities with the PyWFS (e.g. Refs. \citenum{2018JATIS...4d9005H, 2022JATIS...8b1502E, 2023arXiv230509805A}), we do not attempt to implement any within this paper. 
If any individual segment were pistoned above or below the rest of the primary by exactly one wavelength, a single-wavelength PyWFS will not see any disruption signal on the detector, meaning our approach presented in this work would not be able to correct this aberration. Additionally, the PyWFS will generate commands to reach a local minimum in the signal generated on the detector. This means that a segment that is pistoned away by more than one-half of a wavelength will actively be commanded toward the full $\lambda$ offset. As of now, this independent CSP approach is designed for monitoring the primary and is not robust against these phase wrapping complications.

Phase wrapping, 
however, is not a significant risk for the current implementation of the PyWFS on the Keck-II telescope, which makes observations in \textit{H}-band \cite{2020JATIS...6c9003B}. In the \textit{H}-band, a segment would need to reach at least 700 nm of peak-to-valley misalignment prior to risking being driven to a $\lambda$ offset, which is a larger magnitude phase aberration than is currently observed on the Keck primary \cite{2018SPIE10700E..1DR, 2022SPIE12182E..09R}. However, this would become a much larger risk in a visible-wavelength AO system \cite{2023AN....34430088K}. 

Another limitation of our CSP approach is in the sensitivity to inconsistencies in where primary lands on the PyWFS detector. Drifts by even a pixel or less could be significantly problematic. Initial investigations have shown that when the pupil shifts on the detector of the PyWFS by 0.5 pixels, the error in command amplitudes increase by over 500\%. This would suggest that optical alignment stability constraints need to be placed on the system. However, other work has presented improved wavefront reconstruction through intentionally shifting the pupil separation of the PyWFS \cite{2022SPIE12185E..1BB}. Further investigation into the mitigation strategies is required.

Another extra consideration to take for the implementation of our CSP approach is the time and methodology involved for building an interaction matrix of the CSP modes on-sky. In simulation, we build the mixed-space CSP interaction matrix by actuating the piston, tip, and tilt of every segment in both the positive and negative direction and taking the difference in the positive and negative responses from a \tcr{static} 
wavefront as the entry into the interaction matrix. However, in on-sky operation, we will be incapable of injecting \tcr{such a static} 
wavefront \tcr{into} 
the primary and will instead need to build the interaction matrix on-sky. Techniques for this are explored in Ref. \citenum{2006SPIE.6272E..20O}. Many adaptive secondary mirror systems have needed to, or plan to, calibrate their AO system on-sky \cite{2012SPIE.8447E..2BP, 2020SPIE11448E..5JA, 2021MNRAS.501.3443L}, so we will be able to take inspiration from these systems in creating strategies to generate interaction matrices on-sky. 

\tcr{Additionally, real primary mirror segments will have static surface errors as a result of polishing imperfections, harness warping, etc\cite{2024SPIE13094E..3FG}. In the case that these errors are large enough to enter the non-linear response range of the wavefront sensor, they have the potential to impact the measurement of the CSP modes that we intend to control. It is beyond the scope of this paper to fully characterize how the non-linear responses of the PyWFS to phase aberrations at different spatial frequencies impact the measurement and control of the CSP, but we can speculate for general behavior. For static segment surface errors that exist in the controllable space of the DM, we can assume that the AO system will remove the resultant phase aberrations in closed-loop operation, and errors that exist at spatial frequencies that would generate aliasing errors could potentially be controlled with the same methods as other sources of aliasing (e.g. upstream field stops). For the surface errors that exist the non-linear regime of the independent space, it is likely that non-linear phase reconstruction methods would be necessary to prevent errors in the measurement of the CSP modes.} 

\tcr{With these segment surface errors, as we have been using the return-to-reference measurement method for the PyWFS in our simulations, we are also innately constrained by the stability of the reference shape of the mirror segments. When segments are exchanged, or their warping properties are changed, the static reference will also need to be updated. In the case that the segment surface errors meaningfully evolve without the PyWFS reference image being correspondingly updated, we expect a decrease in the accuracy of the iCSP measurements.}

Though this work has primarily focused on applications for the Keck telescopes, the CSP method can be applied to any segmented telescope with an AO system. Notably, applications for the Thirty Meter Telescope and European Extremely Large Telescope have been initially discussed in Ref. \citenum{2023aoel.confE..85C}.  
However, it is worth noting that these telescopes will have approximately 100 times as many segments as the Keck telescopes, and so may need substantially more time dedicated to building an interaction matrix in the presence of an atmosphere unless time-saving strategies are implemented. In our work discussing potential implementation strategies of the CSP method, we additionally investigated methods for optimizing the process of building these matrices \cite{2023SPIE12680E..0CC}, though much work remains to be done with regard to the TMT and E-ELT cases. 

Additionally, we note non-optimal behavior of the CSP when we build a CSP interaction matrix from an aberrated primary. If the primary is not aligned during the process of building the interaction matrix, the resulting command matrix has difficulty sensing various iCSP \tcr{signals,} 
leading to slow modal control. We speculate that the magnitude of the P-V aberrations in the initial misalignments leave the linear range of the PyWFS response curves, leading to \tcr{the optical gain at these modes to generate} 
low-sensitivity command \tcr{responses.} 
It is possible that by building the interaction matrix directly following an external mirror phasing calibration, we can ensure that the primary is aligned to within some requirement for the CSP system to function optimally. Further investigation into this phenomenon will be required prior to attempting to implement this on-sky.

Finally, we conclude our discussion by highlighting that \tcr{this paper only serves as an initial investigation into the potential uses of this methodology. With a sufficiently bright guide star, this methodology could be used in segmented space telescopes to control primary mirror phase to arbitrary phase requirements. As we make no assumptions on the architecture of the wavefront sensor, primary mirror phase may also be controllable with other common wavefront sensors (e.g. non-modulated or triangular pyramids, Shack-Hartmann, Curvature, ZWFS, etc.).} 
There are potentially many sources of slowly evolving wavefront aberrations that could be controlled through active optics on slower timescales than the AO system, and the applications for this sensing method likely go beyond the scope of the authors' original goals.

\section{Key Takeaways} \label{sec:conclusion}
With this project, we have demonstrated the potential for using information from the wavefront sensor that is not used by the rest of the AO system to continue to improve the quality of the wavefront. By simultaneously monitoring the alignment of the primary with the pyramid wavefront sensor, we can substantially decrease the rms surface error of the primary while working in concert with the current edge-sensing method of maintaining the primary's alignment. Between this paper and from part one, we have:
\begin{enumerate}
    \item developed a methodology for identifying and removing the controllable degrees of freedom of individual optics from the signal of a wavefront sensor, 
    \item shown that the independent signal can still be used to measure nearly every CSP degree of freedom to within a precision of \tcr{16\,nm on a $m_H \leq 8$ star} 
    with 30 seconds of effective integration time,
    \item shown that, in the absence of atmospheric turbulence, these measurements effectively eliminate all phase errors from the CSP in the pupil plane, which increases the maximum Strehl ratio achievable by the AO system, 
    \item shown that, despite the presence of atmospheric turbulence limiting the accuracy of the reconstruction of the CSP phase, effective use of the CSP control parameters can mitigate these errors, 
    \item and analyzed some initial on-sky PyWFS telemetry data that validates our approach in using the AO wavefront sensor to independently control the segmented primary mirror. 
\end{enumerate}


\subsection*{Disclosures}
The authors have no relevant financial interests or other potential conflicts of interest to disclose. 

\subsection* {Code, Data, and Materials Availability} 
The data utilized in this study were obtained from W. M. Keck Observatory. Data are available from the authors upon request, and with permission from W. M. Keck Observatory. Code to recreate simulations can be made available upon request to the corresponding author.


\subsection* {Acknowledgments}
The W. M. Keck Observatory is operated as a scientific partnership among the California Institute of Technology, the University of California, and the National Aeronautics and Space Administration. The Observatory was made possible by the generous financial support of the W. M. Keck Foundation. The authors wish to recognize and acknowledge the very significant cultural role and reverence that the summit of Maunakea has always had within the indigenous Hawaiian community. We are most fortunate to have the opportunity to conduct observations from this mountain.
This work was supported by the Heising-Simons Foundation through grant \#2020-1821.


\bibliography{CSP_pt2}   
\bibliographystyle{spiejour}   





\end{document}